\documentstyle[12pt,epsf]{article}
\font\msym=msbm10
\begin{document}
\thispagestyle{empty}
\rightline{KIAS-P00072}
\rightline{UOSTP-00107}
\rightline{{\tt hep-th/0011099}}

\

\def\tr{{\rm tr}\,} \newcommand{\beq}{\begin{equation}}
\newcommand{\eeq}{\end{equation}} \newcommand{\beqn}{\begin{eqnarray}}
\newcommand{\eeqn}{\end{eqnarray}} \newcommand{\bde}{{\bf e}}
\newcommand{\balpha}{{\mbox{\boldmath $\alpha$}}}
\newcommand{\bsalpha}{{\mbox{\boldmath $\scriptstyle\alpha$}}}
\newcommand{\betabf}{{\mbox{\boldmath $\beta$}}}
\newcommand{\bgamma}{{\mbox{\boldmath $\gamma$}}}
\newcommand{\bbeta}{{\mbox{\boldmath $\scriptstyle\beta$}}}
\newcommand{\lambdabf}{{\mbox{\boldmath $\lambda$}}}
\newcommand{\bphi}{{\mbox{\boldmath $\phi$}}}
\newcommand{\bslambda}{{\mbox{\boldmath $\scriptstyle\lambda$}}}
\newcommand{\ggg}{{\boldmath \gamma}} \newcommand{\ddd}{{\boldmath
\delta}} \newcommand{\mmm}{{\boldmath \mu}}
\newcommand{\nnn}{{\boldmath \nu}}
\newcommand{\diag}{{\rm diag}}
\newcommand{\bra}[1]{\langle {#1}|}
\newcommand{\ket}[1]{|{#1}\rangle}
\newcommand{\sn}{{\rm sn}}
\newcommand{\cn}{{\rm cn}}
\newcommand{\dn}{{\rm dn}}
\newcommand{\tA}{{\tilde{A}}}
\newcommand{\tphi}{{\tilde\phi}}
\newcommand{\bpartial}{{\bar\partial}}
\newcommand{\br}{{{\bf r}}}
\newcommand{\bx}{{{\bf x}}}
\newcommand{\bk}{{{\bf k}}}
\newcommand{\bq}{{{\bf q}}}
\newcommand{\bQ}{{{\bf Q}}}
\newcommand{\bp}{{{\bf p}}}
\newcommand{\bP}{{{\bf P}}}
\newcommand{\thet}{{{\theta}}}
\newcommand{\tauu}{{{\tau}}}
\renewcommand{\thefootnote}{\fnsymbol{footnote}}
\

\vskip 0cm
\centerline{ \Large
\bf Noncommutative Vortex Solitons %and Their Low Energy Dynamics
}
%\centerline{ \Large\bf
%of 
% }
\vskip .2cm

\vskip 1.2cm
\centerline{ 
Dongsu Bak,$^a$\footnote{Electronic Mail: dsbak@mach.uos.ac.kr} 
Kimyeong Lee $^{b}$\footnote{Electronic Mail: 
klee@kias.re.kr} and Jeong-Hyuck Park
$^{b}$\footnote{Electronic Mail: 
jhp@kias.re.kr} 
}
\vskip 10mm 
\centerline{ \it $^a$ Physics Department, 
University of Seoul, Seoul 130-743, Korea} 
\vskip 3mm 
\centerline{ \it $^b$ School of Physics, Korea Institute for Advanced 
Study} 
\centerline{ \it 207-43, Cheongryangryi-Dong, Dongdaemun-Gu, Seoul 
130-012, Korea 
} 
\vskip 1.2cm 

%\vskip 1.2cm

\begin{quote}
{%\baselineskip 16pt 
We consider the noncommutative Abelian-Higgs theory
and investigate general static vortex configurations
including recently found exact multi-vortex solutions.
%as well as the self-dual BPS branch of the solutions.
In particular, we prove that the self-dual BPS solutions 
cease to exist once the noncommutativity scale exceeds a 
critical value.
We then study the fluctuation 
spectra about the static configuration and show that the exact 
non BPS solutions are unstable below the critical value.
We have identified the tachyonic degrees as well as massless
moduli degrees. We then discuss the physical meaning of 
the moduli degrees and construct exact time-dependent  
vortex configurations where
each vortex moves independently. 
We finally give the moduli description of the vortices 
and show that the matrix nature of moduli coordinates
naturally emerges.
} 
\end{quote}
%\centerline{\today}

%\pacs{14.80.Hv,11.27.+d,14.40.-n}

\newpage
\section{Introduction}

The noncommutative solitons found in the
noncommutative scalar theory\cite{strominger} do not 
even exist in the commutative version of the 
theory. This indicates that the characteristic 
properties of 
solitons in some  noncommutative field theories 
may greatly
differ from those of ordinary solitons. 
Of course, there are examples where the nature of 
noncommutative solitons and the corresponding
ordinary solitons are quite similar to each other
%the
%corresponding ordinary solitons 
in the sense that
the properties of noncommutative solitons are 
given by just smooth deformation governed by
the noncommutativity scale $\theta$.   

One such example is the U(2) BPS monopole
discussed in Ref. \cite{ahashimoto,bak,hata,grossb}.
The energy and the charge of the BPS monopole 
do not depend on the 
noncommutativity scale.  The effect of the noncommutativity
appears as tilting of D-strings in the transverse space
giving dipole nature of the magnetic charge distribution.
It can be argued that the interactions of the U(2) BPS 
monopoles are independent of the noncommutativity 
scale $\theta$ within the moduli space description of
their dynamics\cite{grossb,yi}.
Contrary to monopole case, the noncommutative scalar solitons
found in \cite{strominger} are genuinely noncommutative 
object since they cannot exist in the ordinary scalar 
theory. As discussed in Ref. \cite{klee}, the shape 
deformation of the scalar soliton is quite peculiar 
when moving
with a constant velocity. Specifically, their 
deformation is 
not simply dictated by the Lorentz contraction but  
described by an area preserving ellipse % shape
exhibiting the UV/IR mixing phenomena of 
noncommutative field theories.
%when an initial configuration has an translational 
%symmetry. 

We here pursue a similar issue on the recently found
exact multi-vortex solutions\cite{dbak} in the noncommutative 
Abelian-Higgs theory\cite{nielsen,jatcar}.
(For soliton solutions of some other models,
see Ref. \cite{gross,polychronakos,
hyang,aganagic,kraus,terashima,khashimoto}.)
Certain apparent properties of the noncommutative 
vortices are  striking even in their static 
properties. 
The multi-vortex solutions are in general not BPS 
saturated states but their energy, nevertheless, scales linearly 
in the number of vortices. This seems to imply 
that there are no interactions between 
vortices even in
this non BPS case. We shall show 
that the self-dual BPS solutions exist only 
when $\theta v^2 \le 1$ where $v$ is the vacuum expectation
value of the Higgs scalar. This property is  also
contrasted to the commutative Abelian-Higgs theory
where the self-dual BPS vortices exist for all  
vacuum expectation values of the scalar.
There is another %rather striking 
aspect concerning
the noncommutative vortex solitons;
the theory allows   exact time-dependent  solutions 
of vortices
each of them moving in an arbitrary velocity from an
arbitrary initial location. In view of generic 
complexity involved with soliton dynamics of field
theory, the existence of such time dependent solutions is
quite peculiar.

In these respects, the systematic approach toward 
the understanding of the noncommutative vortex 
solutions seems imminent on the following 
issues. First  the possible static solitonic 
configurations
need to be mapped out including the self-dual or 
anti-self-dual
BPS branches. Second the stability of the non BPS
multi-vortices is a priori unclear. This issue
can be studied by turning on  general perturbations 
around the static solutions. In case there are tachyonic 
degrees
possessing  a negative mass squared, 
the static configurations are necessarily unstable.
Any small perturbations in this direction will make
the vortices collapse to a  stable 
configuration. On the other hand, when 
fluctuation spectra 
do not possess any tachyonic degrees, any individual  vortex works
as an stable solitonic object. The massless fluctuation is 
responsible for the moduli motions.
Finally, one is interested in the interactions between 
vortices especially when they are stable. The interaction 
can be studied by adopting the scheme of the moduli
space approximation. In fact, one may go beyond the moduli 
space dynamics by identifying quartic 
potential depending on the moduli coordinates in our 
present problem. 
Denoting 
number of vortices by $m$,   
the U($m$) matrix nature of the moduli 
coordinates emerges and the dynamics 
turn out to be described by 
the matrix model of m D0-branes.

In this note, we review first the exact solutions of non-BPS 
multi-vortices.  We also describe the exact solutions 
where vortices are positioned in arbitrary locations.
In Section 3, we  study other static solutions 
focused on
the self-dual BPS branch. 
Our study will be summarized
in Figure 1 where the anti-self-dual branch discussed in 
Ref. \cite{jatcar} is also included.
In Section 4, we study the general fluctuation spectra around 
the static solutions identifying all the tachyonic modes
and massless modes. Masses of the degrees connecting
the vortex to the vacuum can 
be identified by diagonalizing the kinetic and quadratic 
potential terms simultaneously.
The remaining degrees will be shown to be equivalent to the 
fluctuation spectra about the vacuum of the
original Abelian-Higgs system.
In Section 5, we identify the moduli parameters appearing
in the exact solutions by analyzing the %effect of 
translation
and the moments (constructed with help of   
{\it covariant position operator}). We then construct exact time 
dependent solutions describing vortices moving 
in arbitrary velocities. The moduli space description is then 
worked out and the relevant metric will be  shown 
to be flat. We then describe how the matrix nature of the
moduli coordinates emerges.
Last section comprises the summary of our results and concluding
remarks.

\section{Exact multi-vortex solutions}
We begin by recapitulating
the properties of the exact multi-vortex solutions
of the noncommutative Abelian-Higgs theory found
in Ref.\cite{dbak}. 
The noncommutative Abelian-Higgs model in 2+1 
dimensions is described by the Lagrangian %\cite{jatcar}
\begin{equation}
L=-{1\over g^2} \!\int\! d^2 x \Bigl( {1\over 4}F_{\mu\nu} * F^{\mu\nu}
+D_\mu\phi * (D^\mu\phi)^\dagger+{\lambda\over 2} 
(\phi *\phi^\dagger\!-\!v^2)^2\Bigr)
\label{lag}
\end{equation} 
where 
\begin{eqnarray} 
F_{\mu\nu}&=&{\partial_\mu} A_\nu\! -\!{\partial_\nu} A_\mu-i
(A_\mu\! *\! A_\nu\!-\!A_\nu\! *\! A_\mu)\nonumber\\
D_\mu \phi &=& \partial_\mu\phi-iA_\mu\!*\!\phi\,.
\end{eqnarray} 
%and our metric convention is $\eta_{\mu\nu}={\rm diag}(-1,1,1)$.
The  
%Moyal product (
$*$-product is defined by
\begin{equation}
f(x)* g(x)\equiv \Bigl(e^{-i{\theta\over 2}\epsilon^{ij}\partial_i 
\partial'_j} f(x) g(x')\Bigr){\Big\vert}_{x=x'}\,,
\label{star}
\end{equation}
where we  take $\theta$ to be positive without loss of generality.
The theory can be equivalently presented by operators on the Hilbert space
defined by 
\begin{eqnarray}
[\hat{x}\,,\,\hat{y}]=-i\theta 
\end{eqnarray}
%$[\hat{x},\hat{y}]=-i\theta$ 
where the $*$-product
between functions becomes the ordinary product between the 
%corresponding 
operators. 
For given 
function 
\begin{eqnarray}
f(x,y)=\int {d^2 k\over (2\pi)^2} \,
\tilde{f}(k) e^{i(k_x x+k_y y)}, 
\end{eqnarray}
%$f(x,y)=\int {d^2 k\over (2\pi)^2} \,
%\tilde{f}(k) e^{i(k_x x+k_y y)}$, 
the corresponding operator can be found
 by
the Weyl-ordered form of
\begin{eqnarray}
\hat{f}(\hat{x},\hat{y})=\int  {d^2 k\over (2\pi)^2} \,
\tilde{f}(k) e^{i(k_x \hat{x}+k_y \hat{y})}. 
\end{eqnarray} 
%$\hat{f}(\hat{x},\hat{y})=\int  {d^2 k\over (2\pi)^2} \,
%\tilde{f}(k) e^{i(k_x \hat{x}+k_y \hat{y})}$. 
One may then easily show that 
%$\hat{f} \hat{g}=\hat{f\!\!*\!\!g}$,
$\int d^2x \,f$ is replaced by 
$2\pi\theta \,\tr \hat{f}$ and $\partial_i f$
corresponds to $-{i\over \theta}\epsilon_{ij}[ \hat{x}_j, \hat{f}]$. 
With the operator-valued fields,
the action can be written as 
\begin{equation}
L=-{2\pi\theta \over g^2} \tr [ \,{1\over 4}F_{\mu\nu} F^{\mu\nu}
+ D_\mu\phi(D^\mu\phi)^\dagger +{\lambda\over 2} (\phi\phi^\dagger-v^2)^2]
\label{lag1}
\end{equation} 
where hats are dropped for simplicity and the derivative notation
%$\partial_i\, f$ 
is understood as  $\partial_i f\equiv 
-{i\over \theta}\epsilon_{ij} [x_j, f]$.

At this point, we introduce the creation and 
annihilation operators by
$c^\dagger\equiv {1\over \sqrt{2\theta}}{(x+iy)}$ and by
$c\equiv {1\over \sqrt{2\theta}} 
{(x-iy)}$, which satisfy $[c,c^\dagger]=1$.  
To represent arbitrary operators in the Hilbert space we
shall use the occupation number basis by $G= \sum
%\sum^\infty_{n=0} 
g_{kl}|k\rangle\langle l|$
with the number operator $%\hat{n}=
c^\dagger c$.
 We will further denote
$A=A_x-iA_y$, $\,\partial_- G\equiv (\partial_x-i\partial_y) G=
\sqrt{\frac{2}{\theta}}[c,G]$
and  $\partial_+ G\equiv (\partial_x+i\partial_y) G=
-\sqrt{\frac{2}{\theta}}[c^\dagger,G]$.

The system is invariant under the gauge transformation,
\begin{eqnarray}
A_\mu'=U^{\dagger} A_\mu U +iU^{\dagger}\partial_\mu U\,, \ \ \
\phi'=U^{\dagger}\phi\,,
\label{gauge}
\end{eqnarray}
where the gauge group element $U$ satisfies 
\begin{eqnarray} 
U U^\dagger=U^\dagger U=I.
\end{eqnarray}
%$U^\dagger U=U U^\dagger=I$.
We introduce a covariant quantity
$K$ defined by
 \begin{eqnarray}
A=-i\displaystyle{\sqrt{\frac{2}{\theta}}}\, (c-K)\,,
\label{definitionofk}
\end{eqnarray}
which transforms %covariantly 
as
%\begin{eqnarray}
$K'= U^{\dagger}K U$ 
%\label{gaugek}
%\end{eqnarray}
 under the gauge 
transformation in (\ref{gauge}). Later it
will be interpreted
as a covariant version of position
operator up to numerical coefficient.

The Hamiltonian can be constructed as
\begin{eqnarray}
H={2\pi\theta \over g^2} \tr 
[ {1\over 2}(E^2+B^2)+D_t\phi (D_t\phi)^\dagger
+ D_i\phi (D_i\phi)^\dagger +{\lambda\over 2} 
(\phi\phi^\dagger-v^2)^2]\,.
\label{energy}
\end{eqnarray}
using the time translational invariance of the system.
On the  gauge choice $A_0=0$,
the equations of motion read
\begin{eqnarray}
&& \,\,\ddot{\phi}-D_iD_i\phi+\lambda  
(\phi\phi^\dagger-v^2)\phi=0\,,\nonumber\\
&&  \ddot{A_i}+\epsilon_{ij}D_jB=J_i\equiv i %\epsilon_{ij}
[\phi (D_i\phi)^\dagger -D_i\phi\, \phi^\dagger]\,,
\label{eqofmotion}
\end{eqnarray}
with the Gauss law constraint
\begin{eqnarray}
D_i\dot{A_i}=J_0\equiv i %\epsilon_{ij}
[\phi \dot\phi^\dagger -\dot\phi\, \phi^\dagger]\,. 
\label{constraint}
\end{eqnarray}

The exact multi-vortex solutions found in Ref. \cite{dbak}
are given by
\begin{eqnarray}
K= S_m c S^\dagger_m\,,\ \ \ \phi=v S_m\,,
\label{unitsolution}
\end{eqnarray}
where $S_m$ denotes the shift operator $S_m=\sum^\infty_{n=0}
|n\!+\!m\rangle  \langle n|\,\, ( m > 0)$.
The shift operator satisfies relations
\begin{eqnarray}
 S^\dagger_m S_m =I\,,\ \ \ S_m S^\dagger_m=\bar{P}_m \equiv I-P_m\,,
\label{relations}
\end{eqnarray}
with the projection operator $P_m$ defined by
\begin{eqnarray}
 P_m=\sum^{m-1}_{a=0}|a\rangle  \langle a|\,.
\label{projection}
\end{eqnarray}
The magnetic field of the solitons reads
\begin{eqnarray}
B={1\over \theta}P_m\,.
\end{eqnarray} 
The flux defined by $\Phi\equiv \theta \tr B$ %of the vortices
is %fiven by 
$m$ on the solution. Thus the solution describes
$m$ vortices of the Abelian-Higgs theory characterized 
by topological 
quantity $\Phi$.
The energy of the vortices is evaluated as
\begin{eqnarray}
M(v,\theta)= {\pi m\over g^2} 
\left({1\over \theta} +\lambda \theta v^4\right)\ge
{2\pi m\over g^2}{\sqrt{\lambda}} v^2\,. 
\label{oneenergy}
\end{eqnarray}  
When $\lambda=1$, the theory allows so called Bogomol'nyi bound as discussed 
in Ref.~\cite{jatcar}. In fact it is straightforward to verify that the 
energy functional can be expressed as a complete squared form plus a
topological term by
\begin{eqnarray}
H={\pi\theta \over g^2} \tr [
%E^2+2(D_0\phi)(D_0\phi)^\dagger
(B\pm (\phi\phi^\dagger-v^2))^2+2(D_\pm\phi)(D_\pm\phi)^\dagger
\pm\epsilon_{ij}D_iJ_j\pm 2 v^2 B] \ge {2\pi v^2\over g^2} |\Phi|\,,
\label{bound}
\end{eqnarray}
where we omitted the kinetic terms involving  $E_i$ 
and $D_t\phi$.
The saturation of the bound occurs once the self-dual
Bogomol'nyi equations,
\begin{eqnarray}
D_+ \phi=0,\ \ \   B=v^2-\phi\phi^\dagger\,, 
\label{bpsselfdual}
\end{eqnarray}
%$D_+\phi=0$ and $B=1-\phi\phi^\dagger$,
or the anti-self-dual equations
\begin{eqnarray}
D_- \phi=0,\ \ \   -B=v^2-\phi\phi^\dagger\,, 
\label{antiselfdual}
\end{eqnarray} 
are satisfied. 
When $\lambda=1$,
the bound in (\ref{oneenergy}) agrees with the Bogomol'nyi bound
that is an absolute energy  bound for $m$ vortex solution. Hence
when $v^2={1\over \theta}$ and $\lambda=1$, the solution should be 
a BPS solution. Indeed for the specific value of $\theta v^2$, 
one can check
that the solution satisfies the self-dual 
%[upper sign of (\ref{bpsequation})] 
BPS equations. 
This BPS solution is clearly  stable because they
saturate the energy bound set by the topological quantity.

Other obvious generalization of the static multi-vortex solution
is given by\cite{terashima}
\begin{eqnarray}
K= S_m c S^\dagger_m +{1\over \sqrt{2\theta}}
\sum^{m-1}_{a=0}\lambda_a
|a\rangle  \langle a|
\,,\ \ \ \phi=v S_m\,,
\label{generalexact}
\end{eqnarray}
where $\lambda_a$'s are constant complex 
numbers. This solution has the same flux and energy as
the solution in (\ref{unitsolution}). Hence we see that
$\lambda_a$ is the moduli parameters of the multi-vortices. 
Later we shall clarify the stability of the vortex solutions,
which is a priori not clear because they are not always
BPS saturated solutions. But before discussing this 
matter, we will study 
the BPS solutions for 
$\lambda=1$ and $\theta v^2 \neq 1$
or other possible static solutions.

\section{BPS solutions of multi-vortices}

In the last section, we have derived the BPS equations of the Abelian-Higgs 
theory with $\lambda=1$. The static multi-vortex solutions are in general 
not BPS saturated. However they become self-dual BPS solutions
for a special value of $\theta v^2=1$.  
%When $\lambda=1$ the Maxwell-Higgs theory has ${\cal N}=2$ 
%supersymmetric extension and there are  BPS bound states. 
%The BPS equations are
%\begin{equation}
%\begin{array}{cc}
%B=v^{2}-\phi\bar{\phi}\,,~~~~&~~~~D_{+}\phi=0\,,
%\end{array}
%\label{BPS}
%\end{equation}
%while the anti-BPS equations are
%\begin{equation}
%\begin{array}{cc}
%B=-v^{2}+\phi\bar{\phi}\,,~~~~&~~~~D_{-}\phi=0\,.
%\end{array}
%\end{equation}
In this section we focus on the BPS solutions. 
Some analysis on the anti-BPS solutions is 
carried out in \cite{jatcar} and the comparison will 
follow at the end of this section.
In terms of $K$ %in eq.(\ref{AcK}) 
the BPS equations become
\begin{equation}
\begin{array}{cc}
\displaystyle{\frac{1}{\theta}}\Bigl(1-[K,{K^\dagger}]\Bigr)=v^{2}-\phi{\phi^\dagger}\,,~~~~&~~~~
\phi {c^\dagger}-{K^\dagger}\phi=0\,.
\end{array}
\end{equation}
By virtue of the explicit 
form of ${c^\dagger}$ the latter can be solved
\begin{equation}
\phi=\frac{1}{\sqrt{\theta}}
%\sum_{m=0}^{\infty}
\sum_{n=0}^{\infty}\,\frac{1}{\sqrt{n!}}\,
%\langle m |
{K^\dagger}^{n}| \phi_0\rangle %\,\, | m\rangle
\langle n|\,,
\label{phisol}
\end{equation}
where $\mid \phi_0\,\rangle=\sqrt{\theta}\, \phi | 0\,\rangle$ is 
an arbitrary constant vector. Substituting 
this expression, the BPS equations are reduced to a single equation
\begin{equation}
\theta v^{2}-1+[K,{K^\dagger}]=
\displaystyle{\sum_{n=0}^{\infty}\,\frac{1}{n!}}
\, {K^\dagger}^{n}| \phi_0\rangle\langle \phi_0| K^{n}\,.
\label{master}
\end{equation}
To solve this equation
%\footnote{We 
%have not found the general solution of this equation.} 
we take an ansatz for $K$ as %to be of the form
\begin{equation}
K=\displaystyle{\sum_{n=0}^{\infty}}\,f_{n}|n\rangle
\langle n\!+\!p|\,,
\label{BPSansatz}
\end{equation}
where $p$ is any positive integer.  
Substituting this expression into Eq.~(\ref{master}), one can show the following:
BPS solutions exist only when $1\geq\theta v^{2}$. 
Furthermore, finite energy or 
flux solutions exist only 
for $p=1$. 
%In this case the flux is 
%quantized.  
Specifically we obtain
\begin{eqnarray}
%\begin{array}{ccl}
K&=&\displaystyle{\sum_{a=1}^{m}}\,\sqrt{a(1-\theta v^{2})}
|a\!-\!1\rangle
\langle\,a\mid\,-\,\displaystyle{\sum_{n=1}^{\infty}}\,k_{n}
\mid n\!+\!m\!-\!1\,\rangle\langle\,n+m\mid\,,\nonumber\\
%{}&{}&{}\\
\phi&=&\displaystyle{\frac{\zeta}{\sqrt{\theta}}}
\left(\,|m\rangle\langle 0|\,+
\,\displaystyle{\sum_{n=1}^{\infty}}\,\frac{\bar{k}_{1}\bar{k}_{2}
\cdots \bar{k}_{n}}{\sqrt{n!}}\,|n\!+\!m\rangle\langle n|\,\right)\,,
%\end{array}
\label{BPSsol}
\end{eqnarray}
where $\zeta\in{\mathop{\hbox{\msym \char  '103}}}$,  $m$ corresponds to 
the flux number which is 
non-negative integer, 
and the sequence, $k_{n},\,n=1,2,\cdots$, 
satisfies the recurrence relation
\begin{eqnarray}
%&& q_{n}\equiv | k_{n}|^{2}-n\,,\ \ \ q_{0}=
%m(1-\theta v^{2})\nonumber\\
%&& 
q_{n+1}+q_{n-1}-2q_{n}=
%q_{n}-q_{n-1}+
{q_n\over n}\,\, (q_{n}-q_{n-1}+\theta v^2)\,,
%+\theta v^{2}=|\zeta|^{2}
%\displaystyle{\prod_{l=1}^{n}}
%\displaystyle{\left(1\!+\!\frac{q_{l}}{l}\right)}\,,
\end{eqnarray}
with $q_{n}\equiv | k_{n}|^{2}-n$. %\,,\ \ \ 
The initial data for the recurrence relation  are
\begin{eqnarray}
 q_{0}=
m(1-\theta v^{2})\,\,\ \ \  q_1=m(1-\theta v^{2})+|\zeta|^2-\theta v^2\,,
\end{eqnarray}
where $\zeta$ is an adjustable parameter.
The magnetic field is given by
\begin{equation}
%\theta 
B=\displaystyle{\sum_{a=0}^{m-1}}\,
%\theta 
v^{2} | a\rangle\langle a|\,+
{1\over \theta}\sum_{n=0}^{\infty}\,(q_{n}-q_{n\!+\!1})
|n+m\rangle\langle n\!+\!m|\,,
\end{equation}
so that  the flux is
\begin{equation}
\Phi=m-\lim_{n\rightarrow\infty}q_{n}\,.
\end{equation}
Eq. (\ref{BPSsol}) satisfies the 
BPS equations for any value of $\zeta$.
Choosing $\zeta\!=\!0$ or $|\zeta|\!=\!\sqrt{\theta v^{2}}$ 
gives plus infinity or minus infinity 
flux solution respectively. Furthermore, if $q_{n}$ converges, 
the converging value must be zero. 
Thus by continuity, there exists 
$\zeta$, $0<|\zeta|\leq \sqrt{\theta v^{2}}$, 
which makes $q_{n}$ converge to zero, and hence  
BPS solutions have finite 
and quantized energy. The appendix contains our proof\footnote{
%We conclude this investigation by 
We here like to mention %mentioning 
that
 we have  also numerically verified that 
the value of $|\zeta|$, which makes the series to converge to zero 
up to
a few hundred terms, approaches to a unique value
for a given value of $\theta v^2 \ \in\{   0.1,0.2, \cdots 0.9\}$ and 
for $m=1$.
}. For $\theta v^2=1$, the choice $|\zeta|=1$ leads to the exact solution 
with $q_n=0$ or $k_n=\sqrt{n}$.  %\newline

We have shown that, within the ansatz taken,
there are no BPS solutions possessing a positive flux for 
$\theta v^2 >1$. Without limiting the 
discussions to the specific form, 
one may
prove that there are indeed no self-dual BPS 
solutions for $\theta v^2 > 1$
as an analytic perturbation of small parameter 
$\omega$ around the $\theta v^2=1$ BPS solution.
For this purpose, we shall take a generic perturbation 
around the solution and expand it 
as a power series of the small parameter 
$\omega\equiv \sqrt{|\epsilon|}$ 
with $\epsilon\equiv
\theta v^2 -1$. 
(The choice $\omega=\epsilon$ will quickly lead to
a contradiction.) 
Namely, we consider the fluctuation 
 around the exact solution as
\begin{eqnarray}
&& \phi= v ( 1 + \varphi)S_m
\nonumber\\
&& K= S_m c {S}^\dagger_m +h
\,, 
\label{expansion}
\end{eqnarray}  
with the expansions,
\begin{eqnarray}
 \varphi=\sum^{\infty}_{l=1} \omega^l \varphi_{(l)}\,, \ \ \ \ 
h= \sum^\infty_{l=1} \omega^l h_{(l)}\
\,, 
\label{expansionb}
\end{eqnarray}
and may show that there are no solutions for $\theta v^2 >1$.  
The proof is relegated to the appendix. 

  One could  also try
an expansion with respect to a parameter $\omega_n$ 
defined by $|\epsilon|^{1/n}$
for arbitrary nonnegative integers. Though a little complicated,
one may show that the conclusion remains unchanged.
Thus there are no solutions of the BPS equations for $\theta v^2 >1$
that can be expanded in a power series of $\omega_n$. 
Here we do not turn on the diagonal entry $\lambda_a$ of 
$K$. As will be explained later, the effect of nonzero 
$\lambda_a$ corresponds to locating 
each vortex at $\lambda_a=\lambda^a_x -i\lambda_y^a$ 
position. Considering the case of one vortex, one can easily turn off this 
value by using  the translation symmetry of the system. Hence our proof above
is strictly applicable to this case. Furthermore,
$m$ vortices are an assembly of individual vortices, one naturally 
expects that the above proof goes through even 
$m$ vortices with generic values of 
$\lambda_a$.

\begin{figure}[tb]
\epsfxsize=6.0in
\centerline{
%\hspace{.8in}
\epsffile{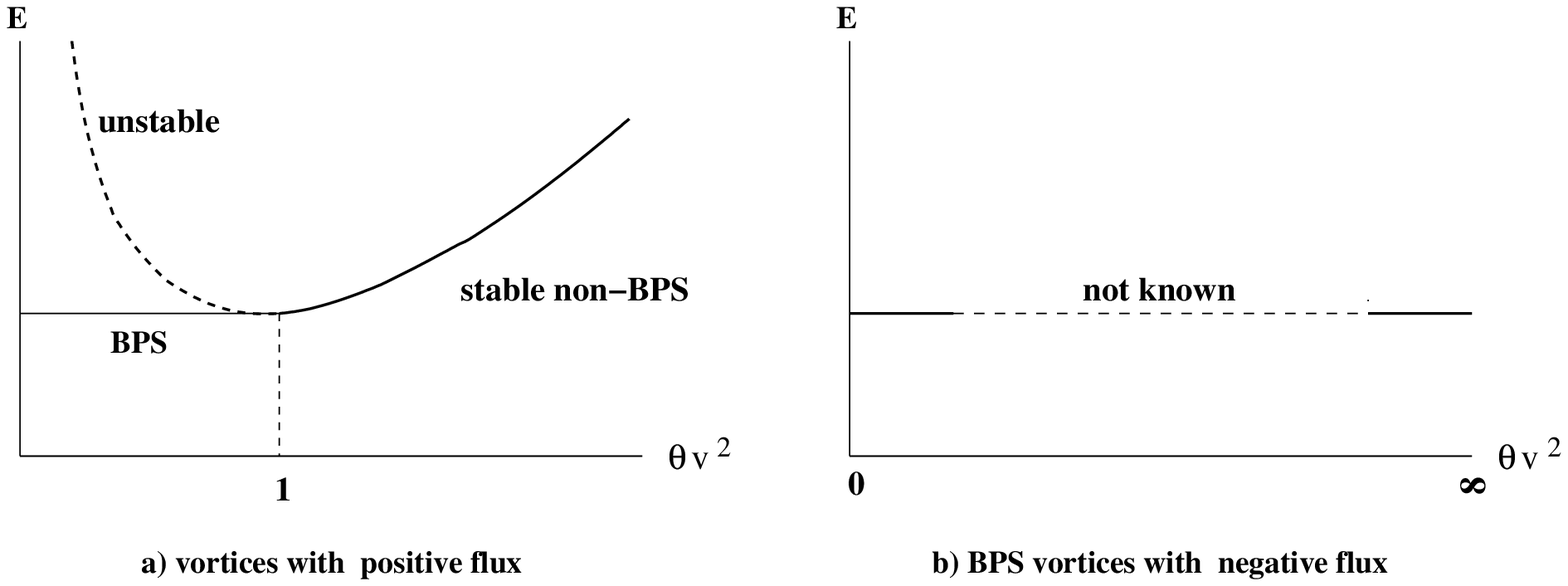}
}
\vspace{.1in}
%\\
 {\small Figure~1:~The energy of (a) the self-dual vortices and (b)
the
anti
 self-dual vortices. We also depict the energy of the non-BPS 
 vortices that have a positive magnetic flux for $\lambda=1$.}
\end{figure}

Figure 1 summarizes our investigation of the static solutions in
the Abelian Higgs theory for $\lambda=1$.
 The self-dual BPS solutions exist only for $\theta v^2 \le 1$.  In the range
the non-BPS exact solutions are unstable due to their higher 
energies.
When $\theta v^2 >1$, the non-BPS branch alone continues 
to exist. For the solutions of a negative flux,
it is shown in Ref. \cite{jatcar} that the anti-self-dual solutions exist
for $\theta v^2 \gg 1$ or $\theta v^2 \ll 1$.
In the intermediate values of $\theta v^2$,
the existence of the self-dual solutions is not known.
The exact vortex solutions with positive magnetic flux exist even for 
$\lambda\neq 1$. It will be shown later that they are also 
stable  only when $\theta v^2 \ge 1$.

\section{Fluctuation spectra around the vortices}
In the last section, we have identified  the possible
static vortex solutions 
including BPS and non-BPS cases. In the BPS case, the classical 
stability of the solution is quite clear because the energy is 
saturating  the bound set by the topological quantity.
For the case of non-BPS, however, it is not a priori clear whether 
the vortices are stable or not. 
When $\theta v^2 < 1$, we have shown that there exist solutions
that have lower energies than the exact non-BPS solutions. 
Thus we expect 
naturally that there should be  tachyonic modes. It is also shown that 
 BPS solutions do not 
exist for $\theta v^2 >1$. Hence in this case the issue of stability 
seems  a different matter. To resolve  these issues clearly, 
we shall
study, in this section, the quadratic fluctuation spectra around
the exact solutions identifying the signature of mass squared for all
possible degrees. It turns out that the solutions $\theta v^2 < 1$
is indeed unstable by developing tachyonic modes in their spectra.
In case of $\theta v^2 = 1$, the potential tachyonic degrees become
massless and the solution is indeed stable. 
For $\theta v^2 >1$, solutions are classically stable 
because the tachyonic degrees become massive. For all these three 
cases, the first $m$ diagonal elements of the gauge field  fluctuation are
massless, which will be identified with 
the degrees of vortex positions.

Let us study first the quadratic fluctuation of  the original 
theory about the vacuum $K=c+{\cal K}$ and $\phi=v(1+h)$ 
without any vortices.
The Lagrangian is then reduced to
\begin{eqnarray}
&& L_v={2\pi \over g^2} {\rm tr}\left[ |\dot{\cal K}|^2 +
\theta v^2 \Bigl(|\dot{h}_R|^2+ |\dot{h}_I|^2 \Bigl)
-{1\over 2\theta} \Bigl|[c, {\cal K}^\dagger ]+[{\cal K}, c^\dagger ]\Bigr|^2
-2v^2 \Bigl|[c, h_R]\Bigr|^2
\right.
\nonumber\\
&& \left. \ \ \ \ \ \  \   -2v^2\Bigl|{\cal K}+i[c, h_I]\Bigr|^2
- 2\lambda \theta v^4 h_R^2
\right]
\,, 
\label{vacuumfluctuation}
\end{eqnarray} 
with the gauss law constraint,
\begin{eqnarray}
[c^\dagger, \dot{\cal K}]+[c,  \dot{{\cal K}}^\dagger]-2i \theta
v^2  \dot{h}_I=0\,,
\label{vacuumgauss}
\end{eqnarray} 
where $h_R\equiv {1\over 2}(h+ h^\dagger)$ and $h_I\equiv {1\over 2i}(h- h^\dagger)$.
One may simplify this action  by reintroducing $A_0$ 
field,  which has a role of imposing the Gauss law 
constraint. We then choose a gauge $A_0= \dot{h}_I$, at 
which ${\cal K} + i[c, h_I]\rightarrow {\cal K}$. The Lagrangian
becomes
\begin{eqnarray}
L={2\pi \over g^2} {\rm tr}\left[ |\dot{{\cal K}}| +
\theta v^2 |\dot{h}_R|^2
-{1\over 2\theta} |[c, {{\cal K}}^\dagger ]+[{\cal K}, c^\dagger ]|^2
%\Bigr.\nonumber\\
%&& \Bigl. \ \ \ \ \ \ \ \ \  
-2v^2 \Bigl(|[c, h_R]|^2
 +|{\cal K}|^2
+\lambda \theta v^2 h_R^2\Bigr)
\right]
\,,
\label{vacuumfluctuationb}
\end{eqnarray} 
with the gauge condition now
\begin{eqnarray}
[c^\dagger, \dot{{\cal K}}]+[c, {\dot{\cal K}}^\dagger]-
i[c^\dagger,[c,h_I]]-i[c,[c^\dagger,h_I]]
-2i \theta
v^2  \dot{h}_I=0\,.
\label{vacuumgaussb}
\end{eqnarray} 
This  can be solved in terms 
of $h_I$ for arbitrary ${\cal K}$,
on which the Lagrangian does not depend.
It is now clear that all the degrees are massive;
the components of ${\cal K}$ have a mass squared greater than 
$2 v^2$ while $h_R$ components a mass squared 
greater than $\lambda v^2$. We see that the gauge field 
absorbs part of the scalar degrees and becomes massive.
This corresponds to the so called Higgs mechanism of the ordinary gauge 
theory when the gauge symmetry is broken spontaneously.

To study the quadratic fluctuation around the exact 
solutions,  we turn on  generic perturbation  of the form
\begin{equation}
\begin{array}{ll}
K =S_m c S_m^\dagger+\Lambda+{\cal  K}\,,\ \ \ \ \phi= v(1+\varphi) S_m\,,
\end{array} 
\label{perturbation}
\end{equation} 
with ${\cal K}$ and $\varphi$ decomposed as
\begin{equation}
\begin{array}{l}
{\cal K}={\cal A}+VS_{m}^{\dagger}+
S_{m}W^{\dagger}+S_{m}\tilde{{\cal K}}S_{m}^{\dagger}=\left(\begin{array}{cc}
{\cal A}&V\\
W^{\dagger}&\tilde{{\cal K}}\end{array}\right)\,,\\
\varphi=XS_{m}^{\dagger}+S_{m}\tilde{h}S_{m}^{\dagger}=
\left(\begin{array}{cc}
0&X\\
0&\tilde{h}\end{array}\right)\,.
\end{array}
\end{equation}
Here we set $\varphi P_m=0$ with out loss of generality 
since an arbitrary $\delta\phi$ can be expressed by $v\varphi S_m$.
%where $\Lambda ={1\over \sqrt{2\theta}}{\rm diag}[\lambda_0,\lambda_1,
%\cdots, \lambda_{m\!-\!1}]$. 
Further introducing unitary operators 
\begin{equation}
U_{a}\equiv\displaystyle{e^{\textstyle{\frac{1}{\sqrt{2\theta}}}
(\bar{\lambda}_{a}c-\lambda_{a}c^{\dagger})}}\,,~~~(0\leq a\leq m-1)\,,
\end{equation}
we parameterize the components deliberately as
\begin{equation}
\begin{array}{ll}
{\cal A}\,\, =\displaystyle{\sum_{a=0}^{m\!-\!1}\sum_{b=0}^{m\!-\!1}}
\,{\cal A}_{ab}\mid a\,\rangle\langle\,b\mid\,,~~~~&~~~~
V=\displaystyle{\sum_{a=0}^{m\!-\!1}\sum_{n=0}^{\infty}}\,V_{an}
\mid a\,\rangle\langle\,n\mid\! U_{a}\,,\\
%{}&{}\\
W=\displaystyle{\sum_{a=0}^{m\!-\!1}\sum_{n=0}^{\infty}}\,W_{an}
\mid a\,\rangle\langle\,n\mid\! U_{a}\,,~~~~&~~~~
X=\displaystyle{\sum_{a=0}^{m\!-\!1}\sum_{n=0}^{\infty}}X_{an}
\mid a\,\rangle\langle\,n\mid\! U_{a}\,,\\
%{}&{}\\
\tilde{{\cal K}}\,=\displaystyle{\sum_{k=0}^{\infty}
\,\,\sum_{n=0}^{\infty}}{\cal K}_{kn}
\mid k\,\rangle\langle\,n\mid \,,~~~~&~~~~
\tilde{h}\,=\displaystyle{\sum_{k=0}^{\infty}\,\,\sum_{n=0}^{\infty}}
\,\tilde{h}_{kn}
\mid k\,\rangle\langle\,n\mid \,.
\end{array}
\end{equation}
The unitary operators, $U_{a}$, satisfy 
\begin{equation}
\begin{array}{ll}
U_{a}cU_{a}^{\dagger}=c+
\textstyle{\frac{1}{\sqrt{2\theta}}}\lambda_{a}
\,,~~~~&~~~~U_{a}c^{\dagger}U_{a}^{\dagger}=
c^{\dagger}+\textstyle{\frac{1}{\sqrt{2\theta}}}\bar{\lambda}_{a}\,,
\end{array}
\end{equation}
which are  helpful in identifying 
variables that diagonalize both the kinetic and potential terms.

Now we insert these to the original Lagrangian in the gauge
$A_0=0$ and expand it to the quadratic terms of the fluctuation. We get
\begin{eqnarray}
&& L_{\rm quad}={2\pi\over g^2}\left[\sum_{ab}\Bigl( 
|\dot{\cal {C}}_{ab}|+ |\dot{\cal G}_{ab}|^2
-{|\lambda_a\!-\!\lambda_b|^2\over \theta^2} |{\cal C}_{ab}|^2
\Bigr) 
+\sum_a \Bigl[|\dot{T}_a|^2 - {\theta v^2\!-\!1\over\theta} |{T}_a|^2 
\Bigr]\right]
\nonumber\\
&&\ \ \ \ \ +{2\pi \over g^2} \sum_{a,n}\left[ |\dot{H}_{an}|^2 +
 |\dot{Y}_{an}|^2 + |\dot{G}_{an}|^2
-{2n\!+\!1\!+\! \theta v^2\over \theta}\Bigl(
 |{H}_{an}|^2 +
 |{Y}_{an}|^2 \Bigr)
\right] + L_{\rm D}
\,, 
\label{quadfluctuation}
\end{eqnarray} 
where $L_{\rm D}$ is same as the Lagrangian (\ref{vacuumfluctuation}) 
but ${\cal K}$ and $h$ and are replaced respectively
by $ \tilde{\cal K}$ and $\tilde{h}$. 
In this Lagrangian, we put
\begin{equation}
\begin{array}{ll}
{\cal C}_{ab}\equiv {1\over \sqrt{2}}[
e^{-i\theta_{ab}} {\cal A}_{ab}-
e^{i\theta_{ab}} {\cal A}^\dagger_{ab}]\,,~~~&~~~
{\cal G}_{ab}\equiv {1\over \sqrt{2}}[
e^{-i\theta_{ab}} {\cal A}_{ab}+
e^{i\theta_{ab}} {\cal A}^\dagger_{ab}]
\,,
\end{array}
\label{definitiondiago}
\end{equation} 
where $\theta_{ab}$ is the argument of $\lambda_a-\lambda_b$ or an arbitrary constant for $\lambda_{a}=\lambda_{b}$.  
We also set  $T_{a}$, $H_{an}$, $Y_{an}$ and $G_{an}$   as
\begin{eqnarray}
&& T_{a}\ \ \equiv \,\, V_{a0}\,,\nonumber\\
&& H_{an}\equiv {\sqrt{\theta v^2}\over \sqrt{\theta v^2\!+\!2n\!+\!1}}
\left(X_{an}\sqrt{2n\!+\!1}- {1\over 
\sqrt{2n\!+\!1}}\Bigl(\sqrt{n\!+\!1}\, V_{a,\, n\!+\!1}+
\sqrt{n}\,
W_{a,\,n\!-\!1}\Bigr) \right)\,,\nonumber\\
&&
 Y_{an}\equiv {1\over \sqrt{ 2n\!+\!1}}
\Bigl(\sqrt{n}\, V_{a,\, n\!+\!1}-\sqrt{n\!+\!1}\,
W_{a,\,n\!-\!1}\Bigr)\,,\nonumber\\
&&
G_{an}\equiv {1\over \sqrt{\theta v^2\!+\!2n\!+\!1}}
\left(\theta v^2 X_{an}+\sqrt{n\!+\!1}\, V_{a,\, n\!+\!1}+\sqrt{n}\,
W_{a,\,n\!-\!1}\right)\,.
\label{offdefinition}
\end{eqnarray}  
To this order, 
the Gauss law constraints for ${\cal A}$ and the off diagonal degrees
become
\begin{eqnarray}
\dot{\cal G}_{ab}=0\  ( {\rm only\  for}\  \lambda_a\neq \lambda_b),
\ \ \ 
\dot{G}_{an}=0
\,,
\label{quadgauss}
\end{eqnarray}   
and, for $L_{\rm D}$, it takes the same form in 
(\ref{vacuumgauss}) where
${\cal K}$ and $h$ %, and $c$ 
are again replaced respectively 
by $ \tilde{\cal K}$ and  $\tilde{h}$. %and $c_m$.

From this it is clear that ${\cal C}_{ab}$ is massless when 
$\lambda_a$ and $\lambda_b$ coincide. 
In particular the diagonal components ${\cal C}_{aa}$
and ${\cal G}_{aa}$ are always
massless;
they are associated with the translational motion of
the vortices. 
The nature of this motion will be exploited
when we discuss the low energy dynamics of the vortices. 
When $\theta v^2
<1$, $T_a$ has a negative mass squared. Hence we see that the 
vortices are unstable even for the case of a 
vortex. On the other hand, for $\theta v^2 \ge 1$, the instability
disappears and the vortex solutions are stable. 
This is also quite consistent with the fact that
there are no BPS solutions 
for $\theta v^2 > 1$. If there were such solutions, 
there must be tachyonic modes because the BPS solution should 
have the lower energy than the non-BPS solutions.

Especially when
$\theta v^2=1$, the potential tachyonic degrees become 
massless and may participate in the low energy dynamics
as will be discussed later.
The remaining of diagonal components are
$H_{an}$ and $Y_{an}$. The $G_{an}$ degrees are dropped out of the
physical space spectrum once the Gauss law constraint is imposed.
Here we were be able to diagonalize these infinite dimensional 
degrees, which is in general not an easy task to achieve.
The spectrum of these physical degrees is particularly simple;
they are all massive with the same 
mass $v^2+{2n+1\over \theta}$, which is independent of the index 
$a$. This spectrum can surely be understood from the underlying
D-brane perspective.

Finally,  $L_{\rm D}$ describes the fluctuation spectra of
the original system around its trivial  vacuum configuration.
This is no coincidence because  the 
degrees of the original system  still remain around vortices. At this point,
we like to emphasize again that they are all massive
controlled by the mass scale $v$ and $\sqrt{\lambda} v$.

\section{Low energy dynamics %of the noncommutative solitons
}

From the analysis of the fluctuation spectra, it is clear
that 
the vortices 
% in (\ref{})
 are unstable due to the  tachyonic modes for 
$\theta v^2 < 1$.
On the other hand,  the vortices  do not exhibit any tachyonic 
instabilities for  $\theta v^2 \ge 1$.    
 For all ranges of parameter $\theta v^2 $,
 the  vortex solutions 
depend upon  $2m$-dimensional free parameters where $m$ is the
topological number corresponds to the total number of 
vortices.  We shall first consider the stable case where 
$\theta v^2 \ge 1$ and begin by clarifying the physical 
interpretation of these parameters. In short, these parameters, 
$\lambda_a$, are positions of vortices on two plane where the 
noncommutative gauge theory is defined. For the gauge group 
element defined by
\begin{equation}
U_P U^\dagger_P =I
\label{matrixgauge}
\end{equation} 
with $ \bar{P}_m U_P  = U_P \bar{P}_m= \bar{P}_m$,  
%($\bar{P}_m\equiv 1-P_m$), 
the corresponding gauge transformation affects only 
first $m\times m$ and $m\times \infty$ component of 
$K$ and $\phi$.  
Utilizing this gauge freedom, we have diagonalized $m\times m$ part of 
$K$ by
\begin{equation}
 P_m K P_m ={1\over \sqrt{2\theta}}\, {\rm diag}
\bigl[\lambda_0,\lambda_1,\cdots,\lambda_{m-1}\bigr]\,.
\label{diagonalmatrix}
\end{equation} 
in the solution (\ref{generalexact}).
%Further utilizing the Weyl subgroup elements, 
Any permutations of the 
eigenvalues $\lambda_a$ and $\lambda_b$ are achieved through
the gauge transformation by the Weyl 
subgroup elements. So they are  physically 
 equivalent configurations.
Thus the moduli
 space is in fact $(R^2)^m/{\cal S}_m$ where ${\cal S}_m$ is the 
permutation group. 

In order to identify the meaning of the moduli parameters, let us first study
the effect caused by the overall translation of the vortex solutions.
For this, we note that the infinitesimal translation is given by
\begin{eqnarray}
&& \delta A_i =-\Bigl(\xi_j\partial_j A_i- D_i (\xi_j A_j)\Bigr)=
B \epsilon_{ij} \xi_j\nonumber\\
&& \delta\phi\  =-\Bigl(\xi_j \partial_j \phi -i(\xi_j A_j)\phi\Bigr)
=-\xi_j D_j\phi\,, 
\label{translation}
\end{eqnarray} 
where we have added the infinitesimal gauge transformation by the
gauge function $\xi_j A_j$. On the solution, this produces
\begin{eqnarray}
 \delta A_i = {1\over\theta}\epsilon_{ij} \xi_j P_m,\ \ \  \delta\phi\  =0\,. 
\label{translationb}
\end{eqnarray} 
The magnetic field $B$ and the Higgs gradient $D_i\phi$ are 
unchanged by the translation
and, consequently,
 one may construct easily the fields translated by a finite amount. Namely, 
the Higgs change is  $\Delta \phi=0$ while 
the change of gauge field in terms of
$K$ variable is given by $\Delta K= {1\over \sqrt{2\theta}}\xi P_m$
where $\xi=\xi_1-i\xi_2$. Hence we see here that the total translation leads 
to a uniform shift of each $\lambda_a$ by the amount $\xi$. This is
of course quite consistent with the interpretation that the moduli parameters
represent positions of the vortices. Of course due to the $U(\infty)$ gauge 
symmetry, the effect of translation does not quite look like a translation of
profile in case of ordinary field theory where a density $\rho(x)$,
for
example, is merely 
shifted by $\xi$ as in $\rho(x-\xi)$ as a result of the translation. 
In this respect whether  
 local informations such as positions of vortices  is well defined in 
noncommutative gauge theory is not obvious at first sight. 
There is another way to get the above result of translation.
The global translation generator can be alternatively expressed as
\begin{eqnarray}
  T= e^{-i \xi_i p_i}\,. 
\label{tgenerator}
\end{eqnarray} 
where $p_i$ is the translation generator $p_i=-{1\over \theta}
[\epsilon_{ij} x_j,\,\,\cdot\,\,]$. In noncommutative field theory,
 the operation
of translation on a field  can be expressed as a similarity 
transformation,
\begin{eqnarray}
  T f(x)= U_T f(x) U^\dagger_T\,. 
\label{tgeneratorb}
\end{eqnarray} 
where $U_T$ is a unitary matrix defined by 
$e^{{1\over \sqrt{2\theta}}(\xi\,
c^\dagger-\bar\xi c)}$.  In our case, we add a gauge 
transformation by $U= U_T$ after the translation. Then the 
resulting gauge and scalar fields read
\begin{eqnarray}
&&  A'=A+{i\over \theta}\xi,\nonumber\\
&&  \phi'=\,\, \phi\,\, U^\dagger_T\,. 
\label{tgauge}
\end{eqnarray} 
The  gauge field  is shifted only by a 
constant piece. We see also  that $\phi\phi^\dagger$ is
invariant. If the scalar were in the adjoint representation,
it would be invariant under the 
transformation. 
%The translation  appears to be almost 
%pure gauge.
In order to obtain the previous result in (\ref{translationb}), we
further perform a gauge transformation by $U=
e^{{1\over\sqrt{2\theta}}(\xi\,
c_m^\dagger -\bar\xi c_m )}$ with 
$c_m\equiv S_m c S^\dagger_m$. 
Using the
explicit expression of the solutions,
one may %then
easily check that results 
agree with (\ref{translationb}). 

One could also study the exact solutions moving in a constant 
velocity as discussed in Ref. \cite{klee}. 
%For simplicity, let us consider the exact solutions of a 
%vortex. 
The theory is not Lorentz invariant because the $*$-product
does not respect the Lorentz symmetry.
However, as discussed in Ref. \cite{klee},
one may still construct moving soliton solutions
once the static solution is given.
The construction is achieved by Lorentz boosting of 
the static solution followed by the change of $\theta$
by $\gamma\theta$ where $\gamma$ is the Lorentz dilation
factor defined by $1/\sqrt{1-\beta^2}$ with velocity 
$\beta$.
Constructed this way, the solution moving in $x$ direction   
reads explicitly, 
\begin{eqnarray}
&& A'_0=-\gamma \beta_x A_x (x',y';\gamma \theta)\,,\ \ \ 
A'_x=\gamma A_x (x',y';\gamma \theta)\nonumber\\
&& A'_y=A_y (x',y';\gamma \theta)\,,\ \ \ \ 
\phi'= \phi (x',y';\gamma \theta)
\,,
\label{lorentz}
\end{eqnarray}  
assuming $A_0=0$ for the static solution.
Here the arguments are given by $x'=\gamma (x-\beta_x t)$ and 
$y'=y$ and  the fields without prime denote any static 
solutions. 
In the present case, one may further simplify
the form of the moving solution again taking 
$A'_0=0$. This gauge choice is achieved from the above
solution by the gauge transformation 
with 
\begin{eqnarray}
U=
e^{-{1\over\sqrt{2\theta}}(\xi\,
\tilde{S}_m \tilde{c}^\dagger 
\tilde{S}^\dagger_m -\bar\xi 
\tilde{S}_m \tilde{c} \tilde{S}^\dagger_m)}
e^{{1\over\sqrt{2\theta}}(\xi\,
\tilde{c}^\dagger 
-\bar\xi 
 \tilde{c})},
\label{movement}
\end{eqnarray}
where we define
\begin{eqnarray}
&& \tilde{c}={\sqrt{\gamma} x-i{y}(\sqrt{\gamma})^{-1}\over 
\sqrt{2\theta}}\,,\nonumber\\
&& \tilde{S}\equiv \sum^\infty_{n=0} 
| n+1\rangle' \langle n|'= \tilde{c}^\dagger
(\tilde{c}\tilde{c}^\dagger)^{-1}
\,,
\label{newcreation}
\end{eqnarray}
and $\xi= \beta_x t$.  
Here  $|n\rangle'$ is the number eigenstate
constructed by the number operator 
$\tilde{c}^\dagger \tilde{c}$.
The form of the solution becomes\footnote{The appearance 
of $c$ instead of $\tilde{c}$ in the gauge field 
is not a typographical 
mistake.}
\begin{eqnarray}
&& 
A'= -i{\sqrt{2}\over \sqrt{\theta}}(c-\tilde{S}_m c \tilde{S}_m^\dagger)
+{i\over \theta}{\beta_x t}\,\, \tilde{P}_m%|0\rangle' \langle 0|' 
\nonumber\\
&& \phi'=\,\,v\, \tilde{S}_m
\,,
\label{movingsolution}
\end{eqnarray}  
with $A'_0= 0$. 
The map from  $c={x-iy\over \sqrt{2\theta}}$ 
to the  new basis 
$\tilde{c}={1\over \sqrt{2\theta}}\left(
x \,{\sqrt{\gamma} -i{y}(\sqrt{\gamma})^{-1}}\right)$ 
belongs to the area preserving 
diffeomorphism. Except some overall numerical 
coefficients, the solution apparently represents
a configuration that has an elliptic 
shape; for example, the magnetic field of a moving
vortex appears in the function representation 
as $2 e^{-{1\over \theta}(x^2\gamma+ y^2 \gamma^{-1})}$.
%One may be tempted to say that such an elliptic shape
%is an physical observable.
%However, it is not because we are dealing with
%a theory with gauge symmetry that includes the $U(\infty)$
%area preserving diffeomorphism.
%In fact
Utilizing the U($\infty$) gauge symmetry, 
the solution (\ref{movingsolution}) can be further
mapped to
\begin{eqnarray}
&& 
A'= -i{\sqrt{2}\over \sqrt{\theta}}(c-{S}_m c {S}_m^\dagger)
+{i\over \theta}{\beta_x t}\,\, P_m%|0\rangle' \langle 0|' 
\nonumber\\
&& \phi'=\,\,v\, {S}_m
\,,
\label{movingsolutionb}
\end{eqnarray}  
by the gauge transformation with the unitary matrix 
\begin{eqnarray}
U_S= \tilde{S}_m S^\dagger_m + \sum^{m-1}_{a=0} |a\rangle' \langle a|
\,.
\label{ss}
\end{eqnarray}  
Inserting (\ref{movingsolutionb}) into the time dependent field equations,
one may directly check that it is indeed 
a solution. Actually, one may even construct solutions 
representing more general motion of  vortices.
The time dependent solutions read
\begin{eqnarray}
&& 
A'= -i{\sqrt{2}\over \sqrt{\theta}}(c-{S}_m c {S}_m^\dagger)
+{i\over \theta}\sum_{a=0}^{m-1}[\lambda^a +\beta^a t]\, 
|a\rangle \langle a| \,\, %|0\rangle' \langle 0|' 
\nonumber\\
&& \phi'=\,\,v\, {S}_m
\,,
\label{movingsolutionc}
\end{eqnarray}  
with $ \beta^a\equiv \beta^a_x-i\beta^a_y $.
The motion of each vortex takes place independently
to an arbitrary direction. 
The magnetic field and electric field are now
\begin{eqnarray} 
B'= {1\over {\theta}}P_m\,,\ \ \  
E'_i=\epsilon_{ij}\sum^{m-1}_{a=0}\beta^a_j |a\rangle \langle a|
\,.
\label{movingfields}
\end{eqnarray}  
The energy of the moving vortices is  evaluated as
\begin{eqnarray}
E(\beta)={1\over 2} 
\left({2\pi\over g^2 \theta}\right)\sum^{m-1}_{a=0}
|\beta_a|^2 -{\pi m\over g^2}\left({1\over \theta}
+\lambda \theta v^4
\right)
\,,
\label{movingenergy}
\end{eqnarray}  
where no approximation is made. The energy behaves precisely
as  free nonrelativistic particles with a mass ${2\pi\over g^2 \theta}$.

One striking fact in the moving solutions 
lies in the fact that there seems no 
limit in the velocity. 
It can apparently exceed the light velocity\footnote{
If the moving solution were not 
exact, we would have easily missed this point.
This is similar to the case of  the noncommutative scalar field theory with 
a quartic interaction. The two particle bound state energy 
is unbounded from  below, 
which was observed in
the exact 
nonperturbative computation of the bound 
state energy\cite{yee}.}. 
On
the other hand, in the original construction by the Lorentz boost followed by
the change in the scale $\theta$, the construction itself loses its validity 
when the velocity exceeds the light velocity. Specifically,
the factor $\gamma$ becomes imaginary. Nonetheless, the final form of the solution 
in this range of velocity   
does solve the time dependent equations of motion.
Our system  lacks the Lorenz invariance and, thus, this seems not a 
serious trouble. But one should remember that the system  originally has
the Lorentz invariance while a specific background field 
(the constant NS-NS two form
background field) is turned on in the string theory 
context\cite{seiberg}.
Without going into details, we like to mention the fact that,
when the velocity exceeds the light velocity,
part of  once stable degrees become tachyonic and instabilities are 
necessarily set in. Hence the solutions seems not to have much physical 
significance when the velocity exceeds the light velocity.
Further investigation is required on this issue.

%which can be visualized investigating
%the magnetic field or energy density.
%(On extracting local informations, see below.)

Let us now turn to the moduli dynamics of vortices.
The study of translation justifies that $\lambda$'s faithfully
represent overall position of vortices. %, but not individual 
%positions. 
%To see this part, 
Let us consider the following 
operator,
\begin{eqnarray}
X_i \equiv x_i-\theta\epsilon_{ij} A_j\,,
\label{position}
\end{eqnarray}  
which may be rewritten equivalently as 
$X=X_1-iX_2= \sqrt{2\theta} K$.
This transforms covariantly under the gauge transformation, i.e.
$X\rightarrow U^\dagger X U$. Since the operator reduces to
$x_i$ in the commutative limit and is gauge covariant, we shall 
call it as {\it covariant position operator}. 
Another justification for the terminology comes as follows.
It transforms as
\begin{eqnarray}
X'_i =X_i+\xi\,
\label{transposition}
\end{eqnarray}
under the translation of (\ref{tgeneratorb}) followed by
the gauge transformation by $U=U_T$. This is precisely 
the required 
property as a {\sl position} operator under translation up to
gauge freedom. 
It will be used to 
measure local properties of the noncommutative field theory.
To show that the eigenvalues $\lambda_a$ represent 
positions of vortices, let us consider the following 
moments
\begin{eqnarray}
I_{k,l} \equiv  2\pi\theta {\rm tr} [X^k (X^\dagger)^l H]\,.
\label{moment}
\end{eqnarray}  
These quantities are  gauge invariant and measure the local
distribution of
matters in noncommutative gauge  theory. For example, $I_{1,1}$
corresponds  to the moment of inertia  for the configurations of 
 the ordinary field
theory. 

For the exact vortex solutions ($\theta v^2 \ge 1$), 
we have
\begin{eqnarray}
I_{k,l} = M_{\rm one} \sum_{a=0}^{m-1} \lambda^k_a{\bar\lambda}^l_a\,.
\label{solutionmoment}
\end{eqnarray}  
In case of commutative field theory limit, the same moments can 
 be
found only  when the Hamiltonian density is sum of delta function
as 
$H({\bf x})=\sum_{a=0}^{m-1}  M_{\rm one} \delta^2({\bf x}-{\bf
 \lambda}_a)$. Thus we show that the relatively local information of
 noncommutative gauge theory can be obtained from the moments
defined above and that the eigenvalues $\lambda_a$ are representing
 the positions of vortices up to the permutation symmetry.
Considering, for example,  vortices located at the origin,
the size information of the vortex configuration can be 
extracted by the moment of inertia. 
%(measured by the covariant
%position operator). 
 The ``size'' 
(measured by the 
covariant position operator) 
is finite for the BPS vortices 
($\theta v^2 < 1$).  In fact it decreases within the BPS branch
as $\theta$ gets larger and becomes zero for the
stable non BPS vortices ($\theta v^2 \ge 1$).

The moduli dynamics of the noncommutative solitons may be 
pursued in a similar manner as solitons in a ordinary 
field theory. As stated before, we shall consider first the case 
where $\theta v^2 \ge 1$. We proceed by giving the time dependence to 
the moduli parameters and adding an appropriate gauge freedom
so that the motion respects the Gauss law constraint. But in our
present case, it is enough to simply give the time 
dependence without adding any gauge degrees because they already
satisfy the Gauss law constraint. Namely we insert
\begin{eqnarray}
 K = \bar{K} (\lambda_a (t))\,,\ \ \  \phi=\bar{\phi} (\lambda_a (t))
\label{ansatz}
\end{eqnarray} 
to the full 
Lagrangian where quantities with bar denote the vortex 
solutions. (This ansatz is quite consistent with
the moving solutions constructed before.)
The resulting effective Lagrangian is given by
\begin{eqnarray}
L_{\rm eff}= -m M_{\rm one} + {\pi\over g^2\theta}\sum^{m-1}_{a=0}
\dot{\lambda_a}\dot{\bar{\lambda}_a}\,.
\label{effective}
\end{eqnarray} 
Consequently,
the moduli space metric on $(R^2)^m/S_m$ is  flat, i.e.
\begin{eqnarray}
ds^2=\sum^{m-1}_{a=0}
d{\lambda_a}\, d{\bar{\lambda}_a}\,.
\label{metric}
\end{eqnarray} 
The inertia mass here is different from the rest mass
but there is no physical reason why these two masses 
agree. 
Not to mention, this effective Lagrangian can be easily 
quantized and wave functions are those of $m$ free nonrelativistic 
bosons with mass ${2\pi\over g^2\theta}$. 

For the present model, one may in fact go beyond the 
moduli space 
description in discussing the relevant low 
energy dynamics. In the previous section, we studied full
 fluctuation spectra around the static solutions.
We find that off diagonal degrees $W_{aj}$,
$V_{aj}\,\, (j\neq 0)$ and $P_m\varphi \bar{P}_m$ 
(with
$\phi=v(1+\varphi)S_m$)
 are 
massive with a mass squared 
$m_j^2={2j+1\over \theta}+ v^2 \,\, (j\ge 0)$.  
Furthermore $\bar{P}_m {\cal  K} \bar{P}_m $ components
have a mass squared at least order of $v^2$.
The real part of  $\bar{P}_m \,\varphi\, \bar{P}_m $ 
has a mass order of $\lambda v^2$ while its imaginary 
part is a gauge degree of freedom that will be absorbed
into the gauge field $\bar{P}_m {\cal K} \bar{P}_m $
by the Higgs mechanism. 
The alternative description of low energy dynamics is
obtained by ignoring all these massive degrees of 
freedom and focusing on all the remaining fluctuations
around the $\lambda_a=0$ solution. 
Namely we only consider the fluctuation of
the gauge field in $m\times m$ sector
 and the potential tachyonic mode defined by
\begin{eqnarray}
{\cal A} &\equiv& P_m {\cal K} P_m\nonumber\\
|\tauu\rangle&\equiv& %\tauu_a |a\rangle =
\sum_{a=0}^{m\! -\! 1}T_{a} |a\rangle\,.
\label{masless}
\end{eqnarray} 
The full  Lagrangian is then reduced to
\begin{eqnarray}
L_{\rm eff}&=& {2\pi\over g^2\theta}\left[
\theta {\rm tr} \dot{\cal A} \dot{\cal A}^\dagger 
- {1\over 2}{\rm tr}[{\cal A},{\cal A}^\dagger]^2
+ 
\theta \Bigl|\,|\dot{\tauu}\rangle\Bigr|^2
%-{1\over 2} {\rm tr}[C,C^\dagger]^2
- {1\over 2} \Bigl(\Bigl|{\cal A}^\dagger|{\tauu}\rangle\Bigr|^2+
\Bigl|{\cal A}|{\tauu}\rangle\Bigr|^2 \Bigr)\right.
\nonumber\\
&&\ \ \ 
\left.
\ \ \ \ -{3\over 2}\langle{\tauu}|[{\cal A},{\cal A}^\dagger]
|{\tauu}\rangle  
- (\theta v^2-1) \Bigl|\,|\tauu\rangle\Bigr|^2 
-  \Bigl|\,|\tauu\rangle\Bigr|^4 
\right] + L_{\rm res}\,,
\label{effmassless}
\end{eqnarray} 
where the Gauss law constraint
\begin{eqnarray}
&&[{\cal A}, \dot{{\cal A}}^\dagger]- [\dot{{\cal A}}, {{\cal A}}^\dagger]+ |\tauu\rangle \langle\dot{\tauu}| -
|\dot{\tauu}\rangle \langle {\tauu}|=0\nonumber\\
&& \dot{{\cal A}}^\dagger |\tauu\rangle - 
{{\cal A}}^\dagger |\dot{\tauu}\rangle =0
\label{effconstraint}
\end{eqnarray} 
is still in effect on the  Lagrangian.
The residual part of the Lagrangian can be organized as follows.
Denoting all the 
remaining  massive modes collectively by $Z_p$,
there are terms of 
$O(Z^2_p)$,
$O({\cal A} Z_p^2)$, $O(\tauu Z_p^2)$, $O(\tauu {\cal A} Z_p)$,
$O(\tauu^2 Z_p)$,  $O(Z^3_p)$
and quartic terms including at least one massive degrees 
$Z_p$. One should note that there are no terms of order 
$O({\cal A}^2 Z_p)$.

When $\theta v^2 > 1$,  the tachyonic modes become
massive too. 
To truncate the Lagrangian consistently, we consider
${\cal A}\sim O(\epsilon)$. The the ${\cal A}^4$ terms  contribute to the 
Lagrangian as $O(\epsilon^4)$. Now if one turns on any massive degrees,
it should be $O(\epsilon^2)$  due to $Z_p^2$ or $\tauu^2$ terms
in order to have a valid approximation of dropping the 
massive degrees. Then the interaction terms between the massive and 
the massless degrees are of higher order, i.e. $O(\epsilon^n)$ with
$n\ge 5$. For example, we see that  the terms of  $O({\cal A} Z_p^2)$ is of 
order $O(\epsilon^5)$. If there were terms of order $O({\cal A}^2 Z_p)$,
these would contribute to the potential as $O(\epsilon^4)$. 
However, there are no such terms as stated previously.
Hence the massive degrees are effectively decoupled
from the massless degrees to the quartic order in the low 
energies. Hence we may consistently drop all the massive 
degrees consistently.
%So the couplings of the gauge fields $C$ to the massive 
%modes are of the form $C ({\rm massive})^2$.   
Ignoring %the tachyonic modes together with all 
all the  massive
modes, 
we are led to
\begin{eqnarray}
L_{\rm eff}= {2\pi\over g^2}\left[
{\rm tr} \dot{{\cal A}} \dot{{\cal A}}^\dagger 
- {1\over 2\theta}{\rm tr}[{\cal A},{\cal A}^\dagger]^2
 \right]
\label{effdzero}
\end{eqnarray} 
with a constraint,
\begin{eqnarray}
[{\cal A}, \dot{{\cal A}}^\dagger]+ 
[ {{\cal A}}^\dagger,\dot{{\cal A}}]=0\,.
\label{effgaugeconstraint}
\end{eqnarray}  
This Lagrangian is precisely the  matrix model, which 
coincides with the bosonic part of an effective Lagrangian
for $m$ D0-branes moving in two dimensional 
target space. %In fact this is not merely a  coincidence. 
%The brane picture behind the 
%Abelian Higgs theory for $\lambda=1$
%will be later described and,   there, it will be shown that
%the vortices are nothing
%but the D0-branes moving on an intersection of higher 
%Dp-branes. 

The vacuum moduli of this effective action is the 
vortex moduli described previously by the 
coordinate $\lambda_a$ on $(R^2)^m/{\cal S}_m$.
We see clearly that the %conifold 
singularity when vortices 
are overlapping is resolved in this 
description. Moreover, the
commutative moduli coordinates are replaced  noncommutative
matrix degrees whose structure is especially relevant
when vortices are nearly coincident.
Hence legitimate 
approach toward the quantization of the low energy
dynamics is also quite clear.

One might ask at this point about the nature of the coordinates 
 of vortex positions.  Since the noncommutative space underlies
in defining the noncommutative field theories, one would also expect
that the noncommutative solitons should see directly  
the noncommutative nature of the underling space
through their forms of interactions. But above description does not 
show directly the noncommutative nature. Namely, the interactions
does not show any particular structure 
depending upon the noncommutativity scale 
$\theta$. Stated again, nothing particular happens at the separation 
$\Delta \lambda \sim \sqrt\theta$.
Nonetheless, the vortex positions are truly described not by c-number
eigenvalues but
by matrices. In this respect,  
the locations of vortices still possess a  
noncommutative nature that is originated from the matrix 
properties. 

Next, we consider the case where $\theta v^2=1$. In this case the 
potential tachyonic modes become massless. But there 
exist a quartic 
contributions, so it is not a moduli degree as defined 
by the configuration space of the constant 
energy. But we include it because its contributions  is of 
the same order of ${\cal A}$ when ${\cal A}$ is 
small. Hence, to study interaction between the massless modes and the
massive modes, we let ${\cal A}$ and $\tauu$ be order of $\epsilon$ as
before. Then $Z_p$ may be allowed to  the order of 
$\epsilon^2$ to have a well defined low energy description.
But this time, there are interaction terms of the forms
$\tauu {\cal A} Z_p$ and $\tauu^2 Z_p$, whose contribution to the Lagrangian 
is $O(\epsilon^4)$. Hence the massless degrees are not 
decoupling from massive degrees. 
One could write down the consistent effective Lagrangian
for this case 
too. 
But it turns out that the effective Lagrangian 
 involves infinite number of massive degrees. Instead of
giving detailed analysis, we  here
briefly comment on the nature of the resulting motion 
involving the potential tachyonic modes.  
First note that one may effectively describe the motion
by ${\cal A}$ and $\tauu$ once all the massive modes are 
integrated out. One may then easily verify that, among $O({\cal A}^2
\tauu^2)$ 
terms,
only the term of $\langle \tauu| [{\cal A},{\cal A}^\dagger]|\tauu\rangle $
remains out of (\ref{effmassless}).
This is quite consistent with the translational invariance
of the underlying system, whose action is 
replacing ${\cal A}$ by ${\cal A}+\xi I_{m\times m}$. 
The motion along $\tauu$-direction is controlled by two 
terms, $\langle \tauu| [{\cal A},{\cal A}^\dagger]|\tauu\rangle $ and
$\tauu^4$. This motion excites other components of the magnetic 
field out of the static solution ${1\over \theta}P_m$, while the flux 
$\Phi$ preserved. Hence the motion represents an oscillatory 
dispersion of magnetic field to other components. If the tachyonic modes 
are small enough, the part of matrix mechanics 
responsible for  the vortex positions is, is little 
affected for fixed energies.

Now we turn to the case 
where $\theta v^2 <1$. In this case the fluctuations 
include the tachyonic modes. Small fluctuation will trigger the vortex to run 
into a more stable lower energy configuration that corresponds to  BPS 
states. As shown previously, the BPS state has the
 same flux as the original 
unstable static configuration. 
Thus during the 
process, the flux should be conserved while the difference
 in energy is eventually
dissipated away. The tachyonic instability is present even 
for the case of single vortex. So it can be 
interpreted as a collapse of each 
individual
vortex to a more stable one, i.e. the BPS state.
The detailed study of the collapse will be quite interesting
in relation with recent discussion of the tachyon 
condensation in string theory. 

\section{Conclusions}

In this note, we have first investigated 
general static soliton solutions in the noncommutative Abelian-Higgs
theory. There are exact  multi-vortex solutions found in 
Ref. \cite{bak} for general values of parameters $\lambda$ and $\theta v^2$. 
These are in general non-BPS except 
$\lambda=\theta v^2=1$. We extend these solutions by finding exact solutions 
describing vortices positioned at arbitrary locations.
We have shown that these solutions are unstable only when $\theta v^2 <1$.
It is therefore expected that lower energy non-BPS  solutions exist for
$\theta v^2 =1$ and $\lambda\neq 1$.  
We confirm this by considering self-dual BPS branch for $\lambda =1$. 
For $\theta v^2 \le 1$, %we have shown that 
the self-dual BPS branch develops, which has an 
lower energy than the exact unstable 
vortices. The BPS branch ended at the point  $\theta v^2$ and 
 there no longer exist BPS 
solutions for $\theta v^2 > 1$. Instead, the exact non-BPS 
configurations become 
stable configurations.
We also illustrated the case of 
anti-self-dual BPS solutions that have an
negative flux\cite{jatcar}. The solutions are shown 
to exist for $\theta v^2 \ll 1$ 
or $\theta v^2 \gg 1$. For the intermediate region, 
the existence 
of the BPS
solutions are not clear yet.

We then discussed the general fluctuation spectra 
around the exact static 
vortices with general moduli parameters $\lambda_a$.
It is shown that there are tachyonic instabilities only 
when $\theta v^2 <1$.
We have identified the massless degrees of freedom and 
masses of all the 
off diagonal degrees.  With help of
the  covariant position operator and
studying translation of vortices, we were able to 
identify the physical meaning
of the moduli parameters; they are positions of the vortices. 
We  were  able to construct exact moving solutions 
of vortices, where each 
vortex is 
moving freely  in a arbitrary constant velocity.
We then show that the 
metric in the moduli space is indeed flat by evaluation 
the low energy effective 
Lagrangian within the moduli space description. In fact 
one may go beyond
the moduli space description in this case by 
identifying quartic order interaction terms of the 
massless degrees of 
freedom. It is nothing but the matrix model of $m$ D0-branes moving
in a two dimensional target space. Thus we have shown 
that the low energy
dynamics are faithfully  described not by positions of individual 
vortices but by
matrices.      
 
The exact time dependent solutions describe vortices 
with constant 
velocity. What is striking in the solution is not that 
vortices are moving 
freely but the velocity is not limited by the light velocity.
The solution exists even for the velocity greater than 
the light velocity.
We argued that the fluctuation becomes tachyonic when the 
velocity exceeds
the light velocity. Therefore, the solution seems not to have
much physical meaning when the velocity exceeds the light velocity.
Further detailed study is required on whether the solution 
in the region 
is consistent with special relativity or not. 
Though the system lacks the Lorentz symmetry, 
the special relativity should be still in effect because one may 
regard the 
system as Lorentz invariant system with a specific background field (a 
constant NS-NS two form
background field in string theory) is turned on.

We expect that our investigations %in this note 
can be %easily
generalized to the N=2 supersymmetric version of the 
noncommutative Abelian-Higgs theory. In particular,
supersymmetries will not be preserved even partially for the sector 
of nonvanishing flux with $\theta v^2 >1$.
We like to finally mention that 
our investigations may be applicable to  
other exact solutions 
recently found\cite{aganagic,kraus,terashima,khashimoto}.

\noindent{\large\bf Acknowledgment} 
%We would like to 
%thank 
%P. Yi and C. Zachos  for
%enlightening discussions and R. Jackiw for pointing out 
% Ref.~\cite{hobart}.
%  D.B. would like to thank H. S.  Yang for
%valuable comments. 
This work is supported in part by KOSEF 1998
Interdisciplinary Research Grant 98-07-02-07-01-5 (DB and KL) and by
UOS Academic Research Grant (DB). 
%K.L. thanks the hospitality in Niels
%Bohr Institute where the part of work is done.

\appendix
%\begin{center}
\section{Self-dual BPS solutions}
%\end{center}
\setcounter{equation}{0}
\renewcommand{\theequation}{A.\arabic{equation}}
Here we demonstrate how to obtain the BPS solutions.
Substituting the 
ansatz for $K$ (\ref{BPSansatz})  into the 
master equation\,(\ref{master}) gives
\begin{eqnarray}
&& \sum_{n=0}^{\infty}\,\Bigl(\theta v^{2}-
1+|f_{n}|^{2}-
|f_{n\!-\!p}|^{2}\Bigr)\,\,|n\rangle\langle n|
\nonumber\\
&& =\displaystyle{{\sum_{i=0}^{\infty}\sum_{j=0}^{\infty}
\sum_{n=0}^{\infty}}\,\frac{s_{i}\bar{s}_{j}}{n!}}\,
\bar{f}_{i}\,\,\bar{f}_{i\!+\!p}\cdots\bar{f}_{i\!+\!(n\!-\!1)p}
f_{j}\,\,f_{j\!+\!p}\cdots 
f_{j\!+\!(n\!-\!1)p} |i\!+\!np\rangle\langle j\!+\!np|\,,
\label{subs1}
\end{eqnarray}
where we set $| \phi_0 \rangle=\sum_{i=0}^{\infty}s_{i}\, | 
i\rangle$ and $f_{j}=0$ for any $j<0$. %\newline
Comparing $\mid 0\,\rangle\langle\, i\mid,~i\geq 1$ 
components of the left and right sides we see 
$0=s_{0}\bar{s}_{i},~i\geq 1$. Now by mathematical 
induction, one can show easily $0=s_{i}\bar{s}_{j},
~i\neq j$. Hence we may put
$\mid \phi_0\,\rangle=\zeta\mid m\,\rangle$ for some 
%$c\in %\Com
%$ 
complex number $\zeta$
and a non-negative integer, $m$. 
This 
simplifies Eq. (\ref{subs1}) as
\begin{eqnarray}
%\begin{array}{l}
&& \sum_{n=0}^{\infty}\,\Bigl(\theta v^{2}-
1+|f_{n}|^{2}-
|f_{n\!-\!p}|^{2}\Bigr)\,\,|n\rangle\langle n|
\nonumber\\
&& =|\zeta|^{2}\displaystyle{\sum_{n=0}^{\infty}
\,\frac{1}{n!}}\,
|f_{m}|^2\,| f_{m\!+\!p}|^2\cdots 
|f_{m\!+\!(n\!-\!1)p}|^{2}
|m\!+\!np\rangle\langle m\!+\!np|\,.
%\end{array}
\label{subs2}
\end{eqnarray}
Hence

\begin{equation}
\theta v^{2}-1+\mid f_{n}\mid^{2}-
\mid f_{n-p}\mid^{2}=0\,,
\end{equation}
for 
$0\leq 
n<m$ or
$m<n,\ n\neq m \ {\rm mod}\  p$, %~\mbox{mod}~p\end{array}\right.\,,
and 
\begin{equation}
%\begin{array}{l}
%\theta v^{2}-1+\mid f_{n}\mid^{2}-
%\mid f_{n-p}\mid^{2}=0\,,~~~~~\mbox{for~}
%\left\{\begin{array}{l}0\leq 
%n<N\\N<n,~n\neq N~\mbox{mod}~p\end{array}\right.\,,\\
%{}\\
\theta v^{2}-1+|f_{m\!+\!np}|^{2}-
|f_{m\!+\!(n-1)p}|^{2}=
\displaystyle{\frac{|\zeta|^{2}}{n!}}
|f_{m}|^{2}\,| f_{m\!+\!p}|^{2}
\cdots|f_{m\!+\!(n\!-\!1)p}|^{2}\,.%\,,~\mbox{for}~0\leq n\,.
%\end{array}
\end{equation}
for $0 \le n$.
With $q_{n}\equiv |f_{m\!+\!(n\!-\!1)p}|^{2}-n-\theta 
v^{2}(p-1)(n-1)$, the magnetic field is expressed as
\begin{equation}
 B=v^2 \sum_n{}^{\prime}\,
|n\rangle\langle n|+{1\over \theta}
\displaystyle{\sum_{n=0}^{\infty}}\,[q_{n}-
q_{n+1}-\theta v^{2}(p-1)]\mid np+m\,\rangle\langle\,np+m\mid\,,
\end{equation}
where $\sum{}^{\prime}$ is the sum over $0\leq n<m$ and 
$m<n,\,n\neq m\,{\mathrm mod}\,p$. 
The flux is then given by
\begin{equation}
\Phi%=2\pi\theta\,\tr\, B=2\pi( 
=m\,\theta v^{2}+q_{0}-
\lim_{n\rightarrow\infty}q_{n}\,,
\end{equation}
where if we write $m=pk+r,\,0\leq k,0\leq r\leq p-1$, 
$q_{0}=k(1-\theta v^{2})+(p-1)\theta v^{2}$. 
In order to have %finite flux or 
a finite energy 
$q_{n}$ ought to converge. %\newline
If $\zeta=0$,  we find that 
$| f_{np+r}|^{2}=(1-\theta v^{2})(n+1),\\
0\leq n,\,0\leq r<p$ and $q_{n}=-\theta v^{2}pn+
\theta v^{2}(p-k-1)+k$. 
For this solution, the energy diverges. %This gives a 
%plus infinite energy solution. 
On the other hand, if $\zeta \neq 0$ then $q_{n}$ 
satisfies the following recurrence relation
\begin{equation}
\frac{q_{n+1}-q_{n}+
p\theta v^{2}}{q_{n}-q_{n-1}+p\theta v^{2}}=
1+(p-1)\theta v^{2}+{1\over n}[q_{n}-(p-1)\theta v^{2}]\,.
\label{recAAp}
\end{equation}
We take the $n\rightarrow \infty$ limit of the above equation  
and 
conclude  that $p=1$ is a necessary condition for 
$q_{n}$ to converge. 
Now for $p=1$ let us assume that 
$\lim q_{n}=\alpha$.   This implies that, for any $\varepsilon>0$, 
there exists large $N$ such that $\alpha-\varepsilon<q_{n}
<\alpha+\varepsilon$ for $n\geq N$.  Eq. (\ref{recAAp}) implies
\begin{equation}
\displaystyle{\frac{\theta v^{2}}{q_{N}-q_{N-1}+\theta v^{2}}
=\prod_{n=N}^{\infty}\Bigl(1+\frac{q_{n}}{n}\Bigr)}\,.
\label{<<}
\end{equation}
Furthermore, we have
\begin{equation}
\displaystyle{\prod_{n=N}^{\infty}\Bigl(1+\frac{\alpha-\varepsilon}{n}
\Bigr)}<
\prod_{n=N}^{\infty}\Bigl(1+\frac{q_{n}}{n}\Bigr)
<\displaystyle{\prod_{n=N}^{\infty}\Bigl(1+\frac{\alpha+\varepsilon}{n}\Bigr)}\,.
\label{<<<}
\end{equation}
However for any $\varrho \neq 0$
\begin{equation}
%\begin{array}{ll}
\displaystyle{\prod_{n=N}^{\infty}\Bigl(1+\frac{\varrho}{n}\Bigr)}
=\displaystyle{{\mathrm exp}\left(\sum_{n=N}^{\infty}
\ln (n+\varrho)-\ln n\right)}\,,
%\\
%{}&{}\\
%{}&\simeq\displaystyle{\frac{M^{M}e^{\beta}}{(M+\beta)^{
%M+\beta}}\lim_{x\rightarrow\infty}
%(x+\beta)^{\beta}}=\left\{\begin{array}{ll}\infty~~&\mathrm{for}~\beta>0\\
%0~~&\mathrm{for}~\beta<0\end{array}\right.\,.
%\end{array}
\end{equation}
which is either infinity or zero depending on the signature of
$\varrho$.
Thus, Eqs. (\ref{<<}-\ref{<<<}) implies that %with suitable choice of 
%$\varepsilon$, 
$\alpha$ must be zero. %\newline

With $p=1$, Eq. (\ref{recAAp}) gives a recurrece relation
\begin{equation}
q_{n+1}-q_{n}=q_{n}-q_{n-1}+{q_{n}\over n}\,\,
(q_{n}-q_{n-1}+\theta v^{2})\,,
\end{equation}
with two initial data, $q_{0}=m(1-\theta v^{2})$ and $q_{1}=
m(1-\theta v^{2})+|\zeta|^{2}-\theta v^{2}$. 
Therefore if $|\zeta|^{2}>\theta v^{2}$, then $q_{n}$ is 
monotonically increasing. As the only possible converging value 
is zero,  it must diverge. In case $|\zeta|=0$, it can be easily solved by
$q_{n}=
m(1-\theta v^{2})-n\theta v^{2}$.
%\\
%\begin{flushright}\textit{Q.E.D.}~~~
%\end{flushright}

\section{Nonexistence of self-dual BPS solutions for $\theta v^2 >1$}
\setcounter{equation}{0}
\renewcommand{\theequation}{A.\arabic{equation}}
We shall work in a gauge $ K_{ij}=0 $
for $i > j$.
The BPS equations can be written
as
\begin{eqnarray}
\varphi  c_m^\dagger &=& h^\dagger \bar{P}_m
+c_m^\dagger  \varphi  \bar{P}_m +h^\dagger
\varphi  \bar{P}_m
\label{expansionbpsa}
\\
 \epsilon P_m 
&=&(1+\epsilon)(\varphi  \bar{P}_m +
\bar{P_m}\varphi^\dagger + \varphi  
\bar{P}_m\varphi^\dagger)-([c_m, h^\dagger]+[h,c_m^\dagger ]
+ [h,h^\dagger])
\,, 
\label{expansionbpsb}
\end{eqnarray}  
where $c_m\equiv S_m c S_m^\dagger$ and $\bar{P}_m=1-P_m$.
The relevant part of the  first order equations in $\omega$ reads
\begin{eqnarray}
&& P_m \varphi_{(1)} \bar{P}_m
=-P_m h^\dagger_{(1)} c_m 
+ P_m h_{(1)} c_m^\dagger 
\nonumber\\
&& 
P_m \varphi_{(1)} c_m^\dagger  =
P_m h^\dagger_{(1)} \bar{P}_m
\,.
\label{expansionbps}
\end{eqnarray}
Since $P_m h^\dagger_{(1)} \bar{P}_m=0$ for our gauge choice,
we find that $\varphi^{(1)}_{am}=h^{(1)}_{a,\,m+1}$,
$\,h^{(1)}_{am}$ can be arbitrary but all the remaining
components should vanish. Now we investigate the second order 
equations obtained from the perturbation equation 
(\ref{expansionbpsb}).
Let us   multiply  $P_m$ to the left and to the right
of the equation at the same time. We obtain
\begin{eqnarray}
&&  -{\epsilon\over |\epsilon|}P_m
+ P_m \varphi_{(1)} \bar{P}_m \varphi^\dagger_{(1)}{P}_m
= [P_m h_{(1)} P_m, P_m h_{(1)}^\dagger P_m]
+  P_m h_{(1)} \bar{P}_m h^\dagger_{(1)}{P}_m
\,.
\label{expansionbpss}
\end{eqnarray}
Using the result of the first order equations and taking trace of 
the above equation, one finds
\begin{eqnarray}
&&  -{m \epsilon\over |\epsilon|} 
= \sum^{m-1}_{a=0} |h^{(1)}_{am}|^2
\,.
\label{expansionbpsd}
\end{eqnarray}
Hence we get a contradiction when $\epsilon > 0$.


\begin{thebibliography}{99}




\bibitem{strominger}
R. Gopakumar, S. Minwalla and A. Strominger,
{\sl Noncommutative Solitons},
JHEP  {\bf 05} (2000) 020, 
hep-th/0003160.
%NONCOMMUTATIVE SOLITONS.



\bibitem{ahashimoto} A. Hashimoto and K. Hashimoto, 
JHEP {\bf 11} (1999) 005,
{\sl Monopoles and Dyons
in Non-Commutative Geometry}, 
{hep-th/9909202}.


\bibitem{bak}
%DEFORMED NAHM EQUATION AND A NONCOMMUTATIVE BPS MONOPOLE.
 D. Bak, 
{\sl Deformed Nahm Equation and a Noncommutative BPS monopole},
Phys. Lett. {B 471} (1999) 149,
 hep-th/9910135.


\bibitem{hata}
K. Hashimoto, H. Hata, and S. Moriyama, 
{\sl Brane Configuration from Monopole Solution in Noncommutative
Super Yang-Mills Theory},   JHEP  {\bf 12} (1999) 021, 
hep-th/9910196.
%; K. Hashimoto and T. Hirayama, 
% hep-th/0002090.





\bibitem{grossb}
%DYNAMICS OF STRINGS IN NONCOMMUTATIVE GAUGE THEORY.
D. J. Gross and N. A. Nekrasov, 
{\sl Dynamics of Strings in Noncommutative Gauge Dynamics},
JHEP {\bf 10} (2000) 021, 
hep-th/0007204.


\bibitem{yi}
K. Lee and P. Yi, 
%QUANTUM SPECTRUM OF INSTANTON SOLITONS 
%IN FIVE-DIMENSIONAL NONCOMMUTATIVE U(N)
%THEORIES.
{\sl Quantum Spectrum of Instanton Solitons in Five-Dimensional
Noncommutative U(N) Theories},
Phys. Rev. {\bf D61} (2000) 125015, 
hep-th/9911186. 

\bibitem{klee}
%ELONGATION OF MOVING NONCOMMUTATIVE SOLITONS.
D. Bak and K. Lee,
{\sl Elongation of Moving Noncommutative Solitons},
 hep-th/0007107.


\bibitem{dbak}
D. Bak,
{\sl Exact Solutions of Multi Vortices and False Vacuum Bubbles
in Noncommutative Abelian Higgs Theories},
hep-th/0008204.



%VORTEX LINE MODELS FOR DUAL STRINGS.
\bibitem{nielsen}
H. B. Nielsen and  P. Olesen, % (Bohr Inst.). 1973. 
 Nucl. Phys. {\bf B61} (1973) 45. 


\bibitem{jatcar}
%NIELSEN-OLESEN VORTICES IN NONCOMMUTATIVE ABELIAN HIGGS MODEL.
%By 
D. P. Jatkar, G. Mandal and S. R. Wadia, 
{\sl Nielsen-Olesen Vortices in Noncommutative Abelian Higgs Model},
 JHEP {\bf 09} (2000) 018, 
hep-th/0007078.



\bibitem{gross}
D. Gross and N. Nekrasov, 
{\sl Monopoles and Strings in Noncommutative Gauge Theories},
JHEP {\bf 07} (2000) 034, 
hep-th/0005204.


\bibitem{polychronakos}
A. P. Polychronakos, 
{\sl Flux Tube Solutions in Noncommutative Field Theories},
 hep-th/0007043.
%FLUX TUBE SOLUTIONS IN NONCOMMUTATIVE GAUGE THEORIES.



\bibitem{hyang}
B.-H. Lee, K. Lee and H.S. Yang,
{\sl CP(N) Models on Noncommutative Plane}, 
hep-th/0007140.
%THE CP(N) MODEL ON NONCOMMUTATIVE PLANE.



\bibitem{aganagic}
%UNSTABLE SOLITONS IN NONCOMMUTATIVE GAUGE THEORY.
M. Aganagic, R. Gopakumar, S. Minwalla and  A. Strominger,
Unstable Solitons In Noncommutative gauge Theory, 
 hep-th/0009142. 


%\bibitem{harvey}
%J. A. Harvey, P. Kraus, F. Larsen, and E. J. Martinec,
%{\sl D-Branes and Strings as Noncommutative Solitons},
%JHEP {\bf 07} (2000) 042, 
% 0005031. 
%D-BRANES AND STRINGS AS NONCOMMUTATIVE SOLITONS.




\bibitem{kraus}
%4) EXACT NONCOMMUTATIVE SOLITONS.
J. A. Harvey, P. Kraus and F. Larsen,
{\sl Exact Noncommutative Solitons},
hep-th/0010060. 



\bibitem{terashima}
%2) ON EXACT NONCOMMUTATIVE BPS SOLITONS.
 M. Hamanaka and  S. Terashima,
{\sl On Exact Noncommutative BPS Solitons},
 hep-th/0010221. 


\bibitem{khashimoto}
%1) FLUXONS AND EXACT BPS SOLITONS IN NONCOMMUTATIVE GAUGE THEORY.
 K. Hashimoto,
{\sl Fluxons and Exact BPS Solitons in Noncommutative Gauge Theory},  
hep-th/0010251. 
%







\bibitem{yee}
D. Bak, S. K. Kim, K.-S. Soh and J. H. Yee,
{\sl  Exact Wave functions in a Noncommutative Field Theory},
Phys. Rev. Lett. {\bf 85} (2000) 3087, 
 hep-th/0005253. %; {\tt hep-th/0006087}.





\bibitem{seiberg}
N. Seiberg and E. Witten, 
{\sl String Theory and 
Noncommutative Geometry}, 
JHEP {\bf 09} (1999) 032, 
hep-th/9908142. 






%2) DP - D(P+4) IN NONCOMMUTATIVE YANG-MILLS.
%By Kazuyuki Furuuchi (KEK, Tsukuba). KEK-TH-720, Oct 2000. 16pp. 
%e-Print Archive: 0010119 
 

%3) SOLITONS IN NONCOMMUTATIVE GAUGE THEORY.
%By David J. Gross (Santa Barbara, ITP), Nikita A. Nekrasov
%e-Print Archive: 0010090 







\end{thebibliography}
\end{document}